%% file: main.tex
\documentclass[twocolumn]{aastex631}
\usepackage{threeparttable}
\usepackage{amsmath}

\newcommand{\oiii}{[O\,{\sc iii}]}

\newcommand{\hb}{H$\beta$}
\newcommand{\ha}{H$\alpha$}
\newcommand{\nii}{[N\,{\sc ii}]}

\newcommand{\greenpea}{{\it Green pea}}

\newcommand{\MBH}{$M_{\rm BH}$}
\newcommand{\Mstar}{$M_{\rm *}$}
\newcommand{\Msun}{$M_{\rm \odot}$}
\newcommand{\betaopt}{$\beta_{\rm opt}$}
\newcommand{\betauv}{$\beta_{\rm UV}$}
\newcommand{\itu}{{\it u}}

\newcommand{\iti}{{\it i}}
\newcommand{\itz}{{\it z}}

\begin{document}

\title{Discovery of Local Analogs to JWST's Little Red Dots }

\author{Ruqiu Lin}
\affiliation{Shanghai Astronomical Observatory, Chinese Academy of Sciences, 80 Nandan Road, Shanghai 200030, China}
\affiliation{School of Astronomy and Space Sciences, University of Chinese Academy of Sciences, No. 19A Yuquan Road, Beijing 100049, China}

\author[0000-0002-9634-2923]{Zhen-Ya Zheng*}
\affiliation{Shanghai Astronomical Observatory, Chinese Academy of Sciences, 80 Nandan Road, Shanghai 200030, China}
\affiliation{School of Astronomy and Space Sciences, University of Chinese Academy of Sciences, No. 19A Yuquan Road, Beijing 100049, China}
\correspondingauthor{Zhen-Ya Zheng}
\email{*email: zhengzy@shao.ac.cn}

\author{Chunyan Jiang}
\affiliation{Shanghai Astronomical Observatory, Chinese Academy of Sciences, 80 Nandan Road, Shanghai 200030, China}

\author{Fang-Ting Yuan}
\affiliation{Shanghai Astronomical Observatory, Chinese Academy of Sciences, 80 Nandan Road, Shanghai 200030, China}

\author{Luis C. Ho}
\affiliation{Kavli Institute for Astronomy and Astrophysics, Peking University, Beijing 100871, China}
\affiliation{Department of Astronomy, School of Physics, Peking University, Beijing 100871, China}

\author{Junxian Wang}
\affiliation{CAS Key laboratory for Research in Galaxies and Cosmology, Department of Astronomy, University of Science and Technology of China, Hefei, Anhui 230026, China}

\author{Linhua Jiang}
\affiliation{Kavli Institute for Astronomy and Astrophysics, Peking University, Beijing 100871, China}
\affiliation{Department of Astronomy, School of Physics, Peking University, Beijing 100871, China}

\author{James E. Rhoads}
\affiliation{Astrophysics Science Division, NASA Goddard Space Flight Center, 8800 Greenbelt Road, Greenbelt, MD 20771, USA}

\author{Sangeeta Malhotra}
\affiliation{Astrophysics Science Division, NASA Goddard Space Flight Center, 8800 Greenbelt Road, Greenbelt, MD 20771, USA}

\author{L. Felipe Barrientos}
\affiliation{Instituto de Astrofísica and Centro de Astroingeniería, Facultad de Física, Pontificia Universidad Católica de Chile, Casilla 306, Santiago 22, Chile}

\author{Isak Wold}
\affiliation{Astrophysics Science Division, Goddard Space Flight Center, Greenbelt, MD 20771, USA}
\affiliation{Department of Physics, The Catholic University of America, Washington, DC 20064, USA}
\affiliation{Center for Research and Exploration in Space Science and Technology, NASA/GSFC, Greenbelt, MD 20771, USA}

\author{Leopoldo Infante}
\affiliation{Institute of Astrophysics and Center for Astroengineering, Pontificia Universidad Catolica de Chile, Santiago 7820436, Chile}

\author{Shuairu Zhu}
\affiliation{Shanghai Astronomical Observatory, Chinese Academy of Sciences, 80 Nandan Road, Shanghai 200030, China}
\affiliation{School of Astronomy and Space Sciences, University of Chinese Academy of Sciences, No. 19A Yuquan Road, Beijing 100049, China}

\author{Xiang Ji}
\affiliation{Shanghai Astronomical Observatory, Chinese Academy of Sciences, 80 Nandan Road, Shanghai 200030, China}

\author{Xiaodan Fu}
\affiliation{Shanghai Astronomical Observatory, Chinese Academy of Sciences, 80 Nandan Road, Shanghai 200030, China}



\begin{abstract}
Recently, the James Webb Space Telescope (JWST) has revealed a new class of high redshift (high-$z$, $z>4$) compact galaxies
which are red in the rest-frame optical and blue in the rest-frame UV as V-shaped spectral energy distributions (SEDs), referred to as ``Little Red Dots" (LRDs). It is very likely that LRDs host obscured broad-line active galactic nuclei (AGNs).
In the meanwhile, \greenpea\ galaxies (GPs), which are compact dwarf galaxies at low redshift, share various similar properties with high redshift star-forming galaxies. Here we aim to find the connection between the LRDs and GPs hosting broad-line AGNs (BLGPs).  With a sample of 19 BLGPs obtained from our previous work, we further identify 7 GPs with V-shaped rest-frame UV-to-optical SEDs that are likely local analogs to LRDs.
These V-shaped BLGPs exhibit faint UV absolute magnitudes and sub-Eddington rates similar to those of LRDs. Three of them occupy a similar region as LRDs in the BPT diagram, suggesting they have similar ionization conditions and gas-phase metallicities to LRDs. These similarities suggest that V-shaped BLGPs can be taken as local analogs of high-redshift LRDs. 
In addition, most (16/19) BLGPs, including 6 V-shaped BLGPs, host over-massive black holes above the local \MBH-\Mstar\ relation, making it the first sample of galaxies hosting over-massive black holes at $z<0.4$. These findings will help us learn more about the formation and co-evolution of early galaxies and black holes.

\end{abstract}



\section{Introduction} \label{sec:intro}
Recent spectroscopic observations with the James Webb Space Telescope (JWST) have revealed a population of red compact galaxies at redshift $z > 3$ \citep{Labbe2023, Akins2023, Barro2024, Kocevski2024}. These ``little red dots'' (LRDs), which show red colors in the rest-frame optical, but blue colors in the rest-frame UV (``V-shape''), are likely to host obscured active galactic nuclei (AGNs). This is suggested either by their spectral energy distribution (SED) analysis or by the high detection rate of broad-line emission in the LRD sample \citep{Kocevski2024, Greene2024}. On the other hand, a subset of AGNs identified with broad \ha\ lines just turn out to be LRDs \citep{Matthee2024}.

While the red optical color of LRDs can be attributed to significant dust attenuation in the AGN, the exact physical mechanisms behind their blue UV color remain debated. The UV SED of LRDs can be modeled with either scattered light from an obscured AGN or emission from a young stellar population \citep{Greene2024}. 
Recent JWST/NIRCam imaging of the rest-frame UV has also revealed host galaxies and off-centered host-galaxy structures in several cases \citep{Kocevski2024, Iani2024, Chen2024}. However, the origin of the UV emission requires further investigation with a larger sample.

Additionally, these high-redshift (high-\itz, $4<z<7$) broad-line AGNs are found to host over-massive BHs compared to the local \MBH-\Mstar\ relation \citep{Maiolino2023, Maiolino2024, Harikane2023, Kokorev2023, Furtak2024, Pacucci2023, Yue2024, Stone2023, Chen2024}.  
They can provide valuable insights into the early phase of supermassive black hole (SMBH) growth and the potential roles these faint AGNs may have played in the early Universe, including their influence on cosmic reionization and on their host galaxies.

In the low-redshift (low-\itz, $z<0.4$) Universe, luminous star-forming galaxies (SFGs) with strong emission lines and compact morphology, particularly Green Pea galaxies (GPs), are commonly regarded as low-\itz\ analogs to early Universe SFGs \citep{Izotov2011, Cardamone2009}. 
Pre-JWST studies have highlighted the similarity between low-\itz\ compact SFGs and high-\itz\ SFGs by comparing various properties, such as oxygen abundances, stellar masses, far-UV absolute magnitudes, star-formation rates, LyC photon production efficiencies, UV continuum slopes, narrow-line ratios, and emission-line equivalent widths \citep[][]{Yang2016, Izotov2015, Izotov2021b}.
Spectroscopic data from HST/COS and FUSE have revealed high escape fractions of both Lyman-continuum (LyC) and Lyman-alpha (Ly$\alpha$) photons in some GPs, providing a hint on how ionizing photons from galaxies contribute to the cosmic reionization \citep{Leitet2013, Borthakur2014, Leitherer2016, Izotov2016a, Izotov2016b, Izotov2018, Izotov2021a, Wang2019, Flury2022, Jaskot2019, Henry2015, Kim2020, Izotov2024}.

In the JWST era, studies have demonstrated the similarities between those $z \sim 8 $ star-forming galaxies and low-$z$ GPs, in terms of gas metallicity, emission line ratios, UV luminosity, and size \citep{Rhoads2023, Schaerer2022, Sun2023, Withers2023, Brinchmann2023, Matthee2023}.
GPs can be used to predict the physical properties of high-redshift galaxies and are important for understanding the evolution of galaxies and the processes of star formation and reionization in the early universe.

Given the similarities of a series of properties between GPs and high-\itz\ galaxies,
we wonder if GPs and LRDs, which are both compact, are similar objects at different redshifts.
Since an important feature of LRDs is the high probability of having broad-line AGNs, we will focus on GPs hosting broad-line AGNs.
Recently, we have constructed a sample of 
59 GPs hosting massive black hole (MBH) candidates with broad \ha\ lines, 
which is selected from a parent sample of 2190 GPs from LAMOST and SDSS spectroscopic surveys (\citealt{Lin2024}, hereafter L24). Over 40 percent of these candidates (25/59) are recognized as AGNs via at least one additional method. Furthermore, this sample has systematically more massive black holes than local dwarf galaxies, appearing similar to high-\itz\ faint AGNs recently found by JWST \citep{Lin2024}. 

In this work, we aim to find the connection between BLGPs and LRDs by searching in our BLGP sample for the most important feature of LRDs, which is the V-shaped UV-to-optical SED.
We describe the sample and the method in Section \ref{sec:datasample}. The results are presented in Section \ref{sec:result}. In Section \ref{sec:discussion}, we discuss the implications of V-shaped BLGPs and the presence of over-massive BHs in the local Universe. We conclude this study in Section~\ref{sec:summary}. 
Throughout this paper, we assume the cosmological parameters of $H_0$ = 70 $\rm km\, s^{-1}\, Mpc^{-1}$, $\Omega_m$ = 0.3, and $\Omega_{\Lambda}$ = 0.7.

\begin{figure}[htb]
    \centering
    \includegraphics[width=0.45\textwidth]{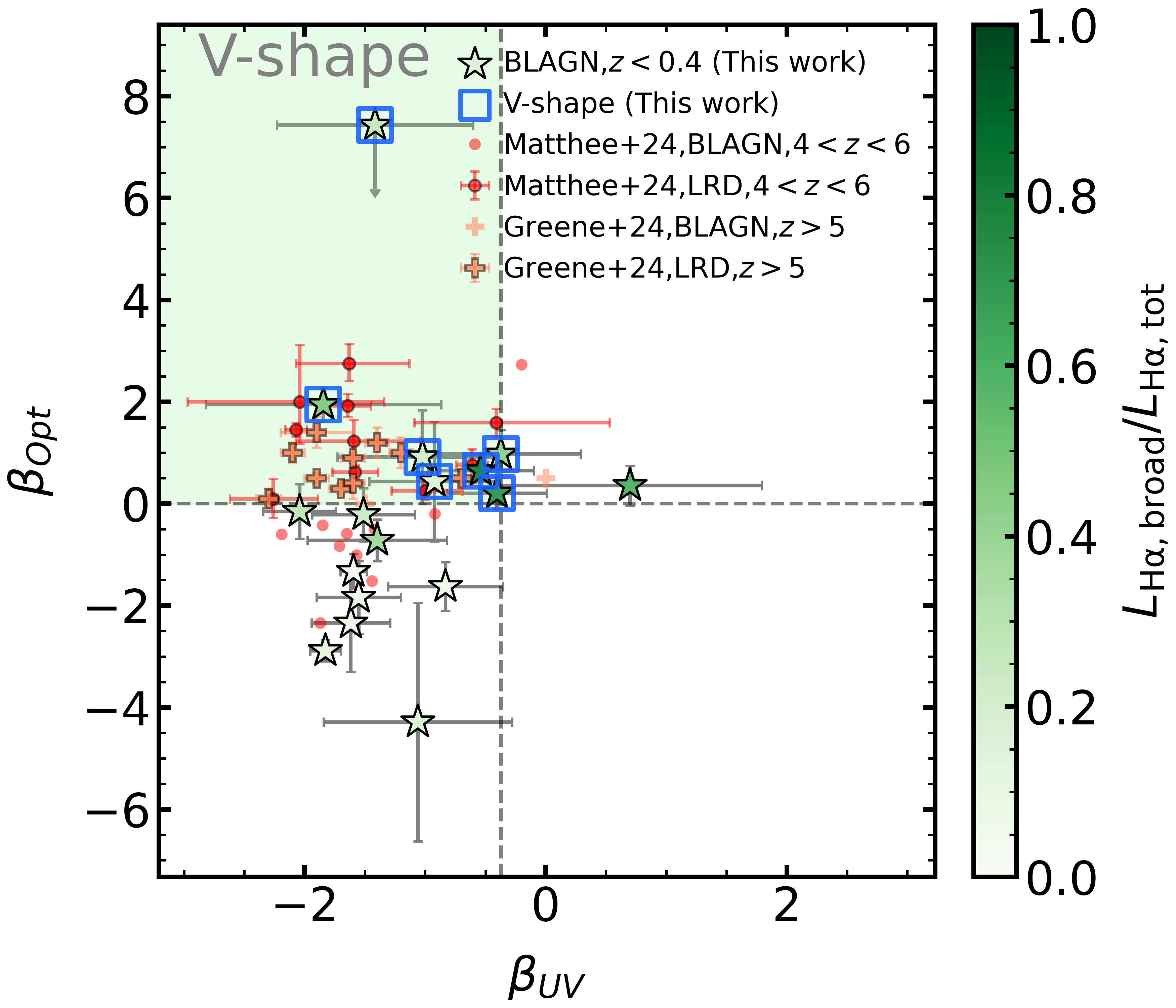}
    \caption{Optical continuum slope \betaopt\ vs. UV continuum slope \betauv\ for the GPs hosting broad-line AGNs (stars). The shaded area presents the LRDs selection criteria \citep{Kocevski2024} with the V-shaped UV-to-optical SEDs. The V-shaped BLGPs are marked on blue squares. The color bar shows the fraction of broad to total \ha\ flux for our BLGP sample. For comparison, we illustrate high-{\it z} broad-line AGNs including LRDs reported in \cite{Matthee2023} and \cite{Greene2024}.}
    \label{fig:slope-slope}
\end{figure}

\begin{figure*}[htb]
    \centering
    \includegraphics[width=0.9\textwidth]{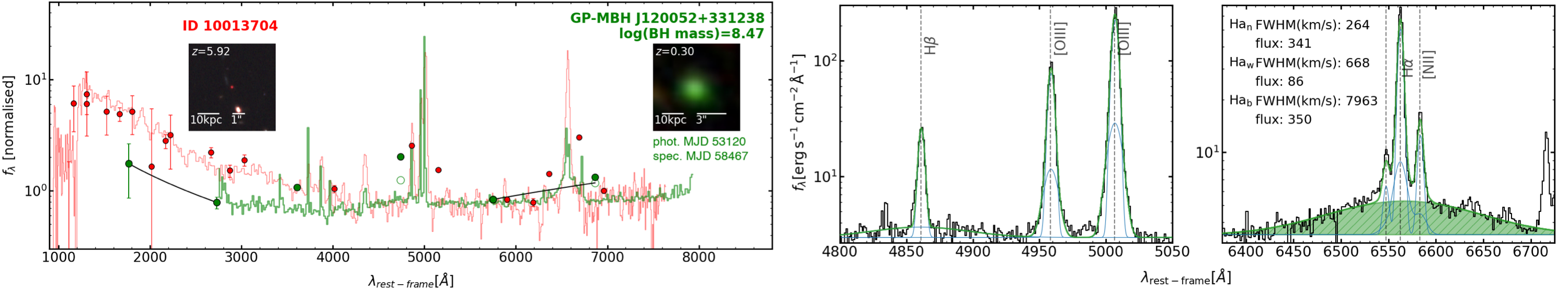}
    \includegraphics[width=0.9\textwidth]{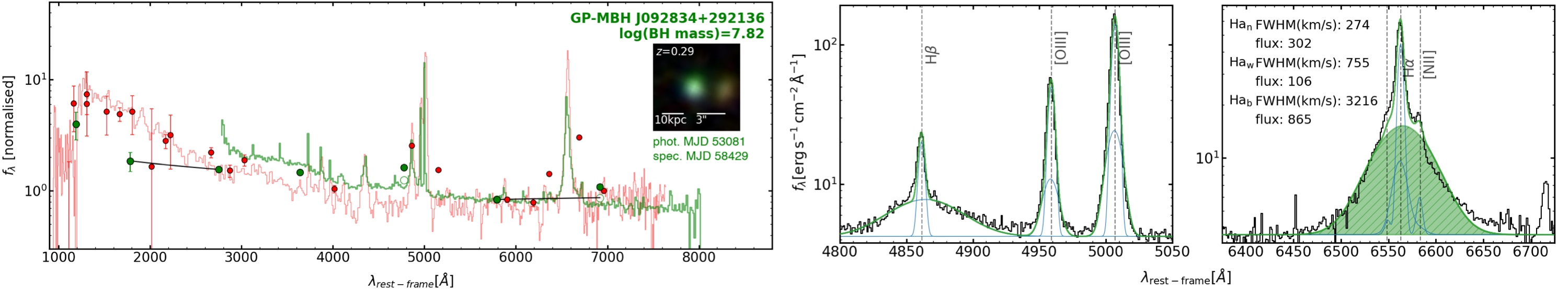}
    \includegraphics[width=0.9\textwidth]{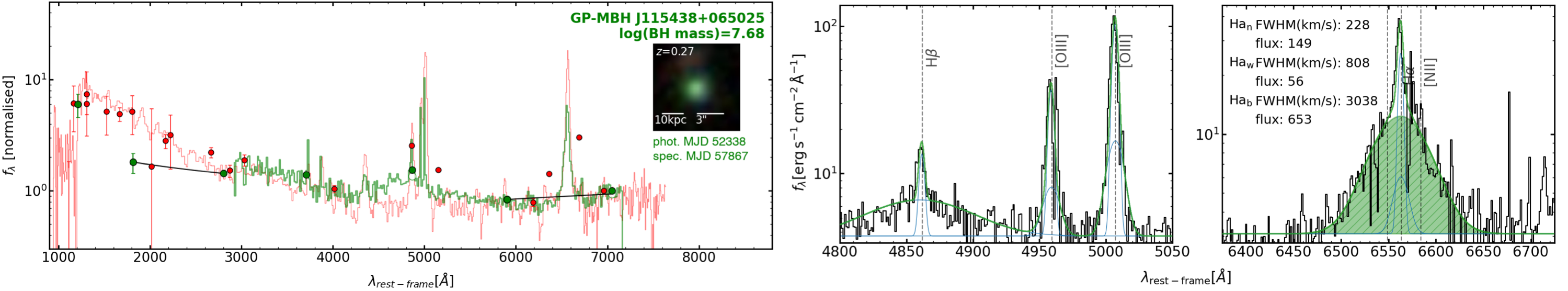}
    \includegraphics[width=0.9\textwidth]{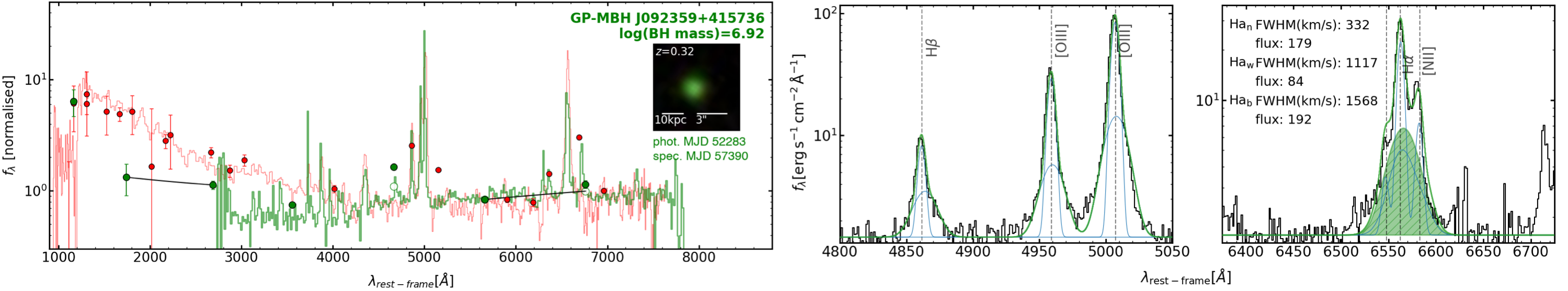}
    \includegraphics[width=0.9\textwidth]{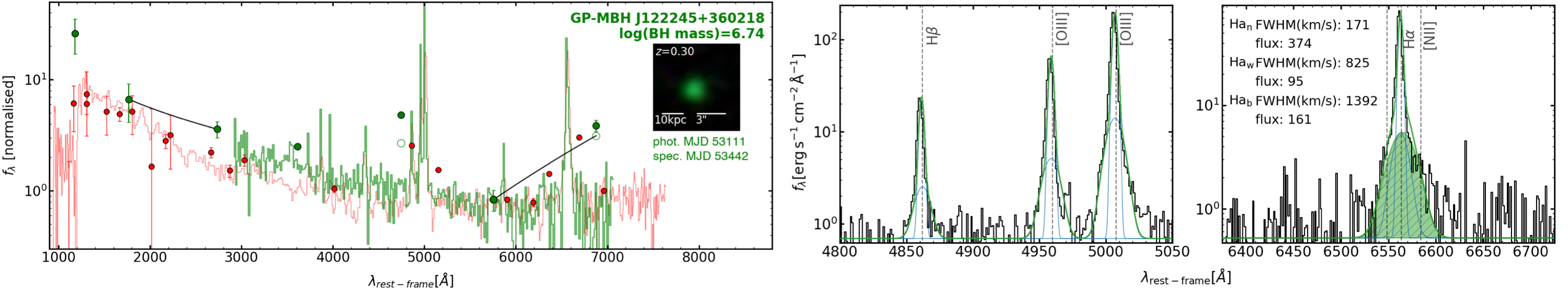}
    \includegraphics[width=0.9\textwidth]{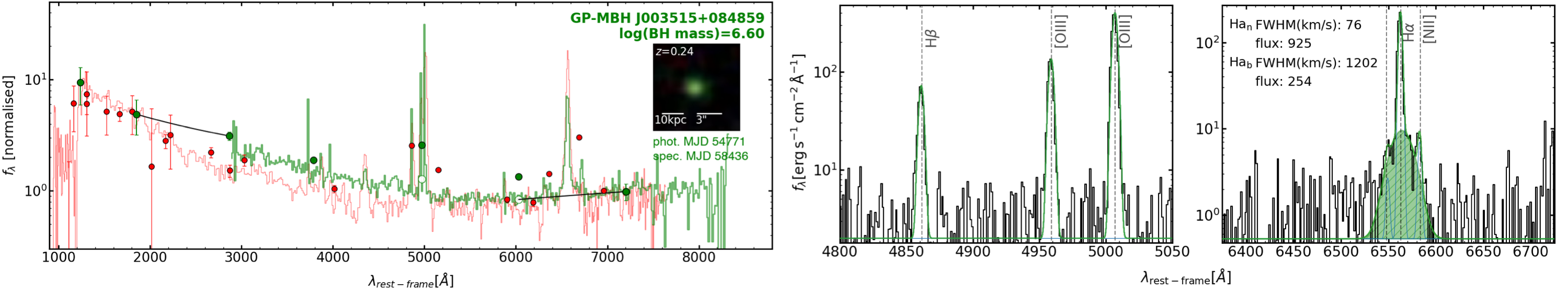}
    \includegraphics[width=0.9\textwidth]{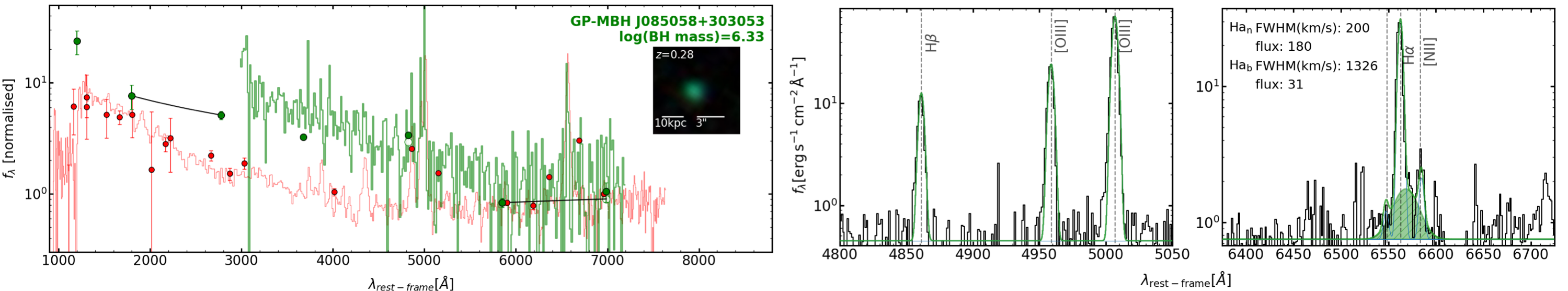}
    \caption{Left column: Comparison of rest-frame UV-to-optical SEDs between the V-shaped BLGPs at $z\sim0.3$ (green dots) and a LRD (JADES 10013704) at $z=5.9$ reported in \cite{Maiolino2023} (red dots). 
    The green circles represent the continuum flux after subtracting emission lines. The background red line and the image in the upper left are the NIRSpec/PRISM spectrum and the NIRCam F115W-F200W-F444W color cutout of JADES 10013704. The green line is the SDSS spectra for each V-shaped BLGP, re-binned to the spectral resolution of NIRSpec/PRISM resolution. All photometries and spectra are normalized at rest-frame 5800 \AA. For each V-shaped BLGP, the SDSS {\it gri} color cutout is also illustrated in the upper right of the corresponding subfigure. The observing dates of the image and spectra for each V-shaped BLGP are also labeled. Right column: Emission-line measurement for the V-shaped BLGPs. The black, blue, and green lines are the observed spectra, the narrow (or plus wide) components, and the composite spectra. The green shade is the broad \ha\ component.}
    \label{fig:sed_spec}
\end{figure*}

\input{tab_para}

\section{Sample and Method}
\label{sec:datasample}
In this section, we describe the sample and the method for selecting the local analogs to JWST's LRDs. The initial sample is the GPs with broad \ha\ lines selected by L24, which is briefly introduced in Section \ref{sec:datasample:blgp}. We then discuss the method of selecting V-shaped BLGPs and their uncertainty in Section \ref{sec:slopemeasurement} and Section \ref{sec:datasample:uncerntainty}, respectively.

\subsection{GPs with Broad \ha\ Lines at $z<0.4$}
\label{sec:datasample:blgp}
In L24, we systematically selected 59 GPs with broad \ha\ lines at $z<0.4$ from a sample of 2190 GPs selected from LAMOST and SDSS spectroscopic surveys. We briefly introduce the selection method and their physical properties below (see L24 for more details).

We search for the broad \ha\ components as the existence of MBHs in GPs. Starting from 2190 GPs' spectra, we first subtracted the stellar continuum and absorption lines to isolate the pure emission-line spectra.
We then fitted \oiii\ lines using either a single narrow component or a combination of narrow and wide components. The narrow component represents the star formation and/or AGNs, while the wide component indicates probable outflows. After subtracting the narrow and possible wide components, we examined the presence of the broad \ha\ component. We found 59 GPs with broad \ha\ components which may host MBHs. 

We adopt the stellar mass \Mstar\ of these GPs from L24. The stellar mass was estimated through the SED fitting with {\tt CIGALE} \citep{Noll2009, Boquien2019, Burgarella2005} on broadband photometry with {\it GALEX}, {\it SDSS}, and {\it WISE}. Compared to L24, we update the measurement of BH mass \MBH\ with the modified relation of virial BH mass in \cite{Ho2015} based on the velocity dispersion and the luminosity of the broad \ha\ component. 
Following \cite{Greene2005}, we use the conversion of velocity dispersion between \hb\ and \ha\ and the relation of \ha\ luminosity and luminosity at the rest-frame 5100 \AA. We adopt the virial coefficient $f$ of 4.47 for \ha\ lines \citep{Woo2015}. This BH mass estimation has a typical uncertainty of $\sim$ 0.4 dex \citep{Zhuang2023}. These inferred parameters are listed in Table~\ref{tab:parameter}.

\subsection{Selecting V-shaped BLGPs}
\label{sec:slopemeasurement}
LRDs have three features, which include broad-line AGNs, compact morphologies, and V-shaped SEDs. First, we require a line width cut of the broad \ha\ line as FWHM$_{\rm H\alpha,\ broad} > 1000\, \rm km\,s^{-1}$, which is the selection criterion for high-\itz\ faint broad-line AGNs in \cite{Matthee2023}. As a result, there are 19 GPs hosting broad-line AGNs (hereafter BLGPs). 

Second, most (14/19) of these BLGPs have compact sizes of {\tt petroRad\_r} $ < 2.5\arcsec$ and appear unresolved in SDSS images, although the spatial resolution of SDSS is much poorer than that of JWST.  

Finally, following the methodology in \cite{Kocevski2024}, we utilize the continuum slopes $\beta$ (where $\beta$ is defined by $f_{\lambda}=\lambda^\beta$) to identify the V-shaped SED in BLGPs. The criteria used here are $\beta_{\rm UV} \lesssim -0.37$ for the blue UV continuum slope and $\beta_{\rm opt} \gtrsim 0$ for the red optical continuum slope, as the shaded region in Figure~\ref{fig:slope-slope}. 
To determine the UV and optical continuum slopes, 
we use photometry from the {\it GALEX} NUV band as well as the SDSS \itu, \iti, and \itz\ cModel photometry, which are presented in Figure \ref{fig:sed_spec}. 
The continuum slope $\beta$ is defined as the following euqation,
\begin{equation}
    \beta = -0.4\frac{m_1-m_2}{\log(\lambda_1/\lambda_2)}-2,
    \label{eq:eq1}
\end{equation}
where the photometric magnitudes (AB) $m_1$ and $m_2$ 
are two continuum bands in the corresponding UV or optical bands.
The measurement errors are calculated as $\sigma_{\beta} = -0.4\cdot\sqrt{( \sigma_{\rm 1}^2+\sigma_{2}^2)}/\log(\lambda_1/\lambda_2)$, where $\lambda_1$ and $\lambda_2$ represent the effective wavelengths of the bandpasses, and $\sigma_1$ and $\sigma_2$ are the photometric errors. 

For the UV slope, to avoid contamination from the strong Ly$\alpha$ line, we use the NUV and \itu\ magnitudes as $m_1$ and $m_2$ in Equation~\ref{eq:eq1}. Out of 19 BLGPs, 17 are covered by {\it GALEX} and have NUV detection.

For the optical slope, to avoid contamination from the extreme \hb-\oiii\ emission-line complex which is difficult to remove from the photometry completely, we use the continuum magnitudes in \iti\ and \itz\ bands as $m_1$ and $m_2$ in Equation~\ref{eq:eq1}, respectively.
Since the \ha\ lines may still contaminate the \iti -band and \itz -band photometry at $0.053\lesssim z\lesssim 0.245$ and $0.245\lesssim z \lesssim 0.411$, respectively, 
we subtract the \ha\ contribution from the photometry to get the continuum magnitudes.
The continuum magnitude $m_{\rm conti}$ is then calculated using the equation $m_{\rm conti} = m_{\rm obs}+2.5\log(\frac{F_{obs}}{F_{obs}-F_{line}})$, where $m_{\rm obs}$ is the observed magnitude and $F_{obs}$ represents the broad-band integral flux converted by $m_{\rm obs}$. The error $\sigma_{m \rm,conti}$ is determined through the error propagation.

\subsection{Uncertainty in Selecting V-shaped BLGPs }
\label{sec:datasample:uncerntainty}
The identification of V-shaped BLGPs in our sample heavily relies on the UV continuum slopes and the optical continuum slopes. The uncertainty in the optical slope is inherently more complex than that of the UV slope due to the influence of extreme optical emission lines. 

The uncertainty in the optical slope arises from the measurement errors in both the emission lines and the continuum. For GP galaxies, the uncertainty in emission-line measurement is generally negligible as these lines are luminous. 
In contrast, the continuum is relatively faint, resulting in a large uncertainty when directly measuring the continuum from the spectra. This is also why we prefer to use the line-subtracted broad-band magnitudes to estimate the optical slopes. 

Another source of uncertainty could be emission-line variability probably caused by supernova explosions, tidal disruption events, or changing-look AGNs \citep{Lin2022}. For instance, the optical slope of J122245+360218 is considered an upper limit because the \itz-band continuum, after subtracting the emission line, appears significantly higher than the corresponding spectra (see Figure~\ref{fig:sed_spec}). The \iti-band magnitude in the spectrum matches the photometric magnitude, indicating no substantial continuum variability but only emission-line variability between the imaging and spectroscopic observing epochs, which were separated by about 330 days.
In addition, this 330-day time lag is insufficient to rule out the possibility of a supernova explosion \citep{Yang2023, Wang2024}.
More recently, the optical variability was also seen in LRDs \citep{Zhang2024}.
Despite these uncertainties, we retain this source as its photometry and optical spectrum profile closely match those of the LRD, JADES 10013704 (see the comparison in Figure~\ref{fig:sed_spec}).

\section{Results} \label{sec:result}
\subsection{V-shaped BLGPs: Local Analogs to LRDs}
Based on the broad \ha\ line width and the continuum slope criteria, we identify 7 V-shaped BLGPs as local analogs to LRDs from a sample of 19 BLGPs.

In Figure~\ref{fig:sed_spec}, we show the rest-frame SEDs and spectra of these V-shaped BLGPs, as well as those of a typical LRD, JADES 10013704, reported in \cite{Maiolino2023}. 
These local analogs exhibit V-shaped UV-to-optical SEDs and strong emission-line features (especially \oiii\ and \ha) similar to those of JADES 10013704. 
We also note a likely inverse correlation between the break wavelength of the V-shaped SED and the BH mass in these V-shaped BLGPs (Table~\ref{tab:v-shapedGP}). This correlation would help us to further constrain many AGN-galaxy hybrid models \citep[e.g.,][]{Greene2024} recently invoked to explain the SED of LRDs.

Additionally, except for V-shaped BLGP with the lowest stellar mass, six others meet the AGN criteria based on the {\it WISE} mid-infrared colors \citep[][also see Table~\ref{tab:v-shapedGP}]{Jarrett2011}. This suggests that they have consistent mid-infrared properties and also align with the scenario of AGNs or starbursts \citep[Figure 12 of][]{Withers2023}.

\begin{figure*}[htb]
    \centering
    \includegraphics[width=0.9\textwidth]{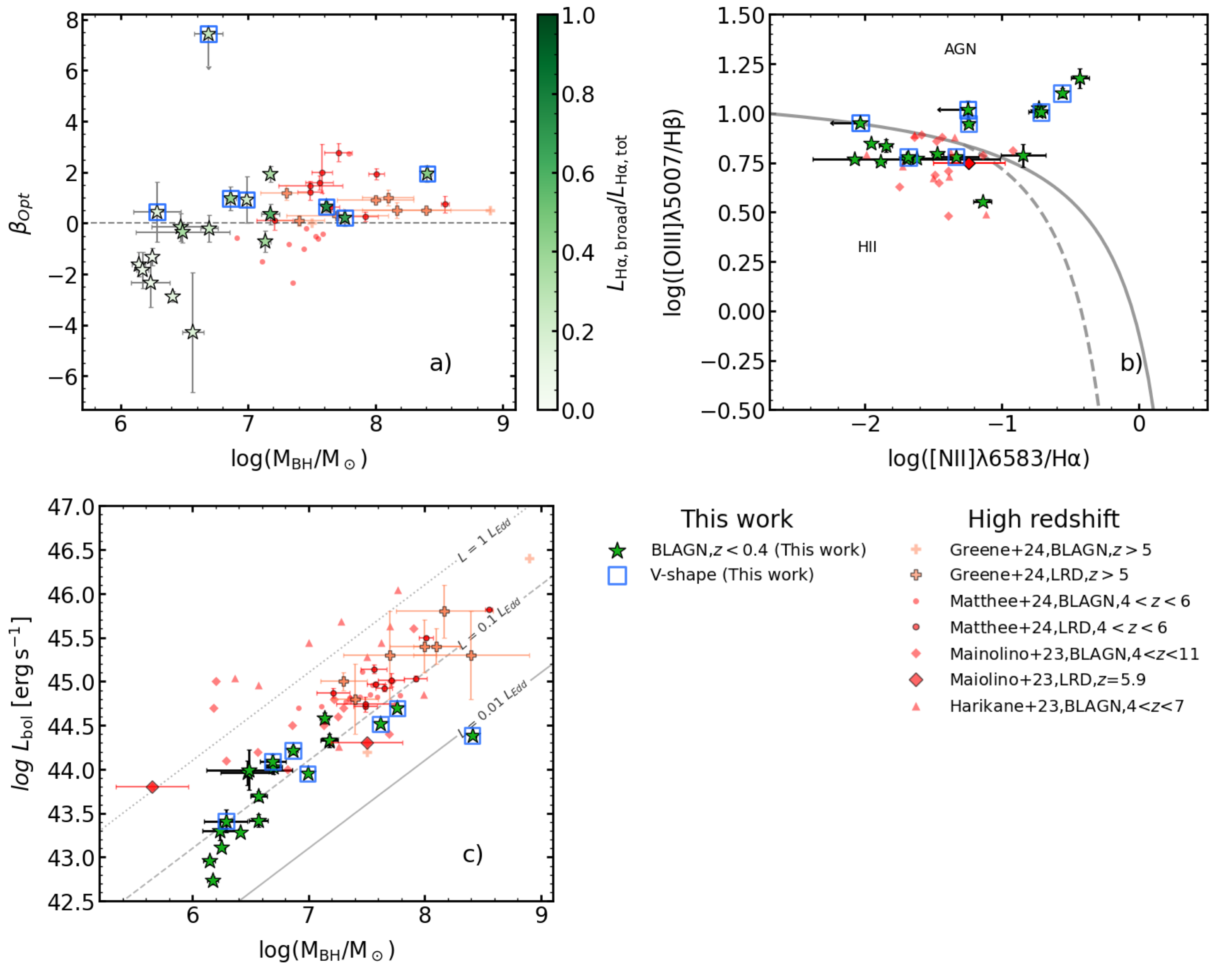}
    \caption{Panel a: Optical continuum slope vs. BH mass for BLGPs (stars).  For comparison, we illustrate high-{\it z} broad-line AGNs including LRDs reported in \cite{Maiolino2023}, \cite{Matthee2023}, \cite{Greene2024}, and \cite{Harikane2023}, if available. 
    Panel b: BPT diagram. The classification lines follow \cite{Kewley2001} (solid line) and \cite{Kauffmann2003} (dashed line), also seen in \cite{Kewley2006}. Here we also mark JADES 10013704 in big diamonds.
    Panel c: Bolometric luminosity vs. BH mass. The grey lines indicate Eddington ratios of $\lambda_{Edd}=$ 1, 0.1, 0.01.
    }
    \label{fig:fig3}
\end{figure*}

\begin{table*}[]
\centering
\caption{Properties of V-shaped BLGPs Ordered by BH Masses} 
\label{tab:v-shapedGP}
\setlength{\tabcolsep}{6mm}{
\begin{tabular}{ccccccc}
\hline\hline
Name& Mid-IR$^a$ & BPT$^a$ & Mg {\sc ii} & Break $\lambda_\textrm{rest}$  & log \MBH & log \Mstar\\
  & AGN & type &  & [\AA]  &[\Msun]& [\Msun]\\
(1) & (2) & (3) & (4) & (5)  & (6) & (7)\\
\hline
J120052+331238 & 1 & AGN & 1 & 2800  & 8.41 & 9.90 \\
J092834+292136 & 1 & AGN & 1 & 2000-3600   & 7.76 & 10.02 \\
J115438+065025 & 1 & AGN & 0 & 2000-3600  & 7.62 & 10.08 \\
J092359+415736 & 1 & AGN & -- & 3000  & 6.86 & 10.03 \\
J122245+360218 & 1 & SF-Seyfert & -- & 4300 & 6.69 & 8.62 \\
J003515+084859 & 1 & SF & -- & 5300  & 6.99 & 8.71 \\
J085058+303053 & 0 & SF-Seyfert & -- & 6000  & 6.29 & 8.33 \\
\hline
\end{tabular}
}
\begin{tablenotes} 
    \item NOTE. Column (1): Galaxy name. Column (2): Classification based on the {\it WISE} mid-infrared color-color diagram \citep{Jarrett2011}, where 1 indicates an AGN and 0 indicates a non-AGN. Column (3): BPT classification.  Column (4): Presence of the Mg {\sc ii}$\lambda 2800$ line, where 1, 0, and -- indicate its presence, the absence, and that it is outside the spectral coverage, respectively. Column (5): Break wavelength at the division between the blue and red sides, which are inspected visually. Column (6): BH mass. Column (7): Galaxy stellar mass. We note that these 7 V-shaped BLGPs are ordered by their BH masses.\\
    $^a$ Classificaition from L24.
    \end{tablenotes}
\end{table*}

\subsection{Similarity Between V-shaped BLGPs and LRDs}
In this section, we present the similarity between V-shaped BLGPs and LRDs in terms of UV obsolute magnitude, the relation of optical slopes and BH masses, narrow-line ratio,  and Edington ratios. 

V-shaped BLGPs have UV absolute magnitudes ranging from -18.9 to -19.8 mag with a median of -19.2 mag. This is similar to those of LRDs \citep{Matthee2024, Maiolino2023, Harikane2023}, placing them in the category of faint AGNs.

Moreover, V-shaped BLGPs have higher \MBH\ and broad-to-total \ha\ luminosity ratio ($L_{\rm H\alpha,broad}/L_{\rm H\alpha,tot}$) compared to non-V-shaped BLGPs (see Figure \ref{fig:fig3}-a). We find a positive correlation between \betaopt, \MBH, and $L_{\rm H\alpha,broad}/L_{\rm H\alpha,tot}$, consistent with the samples of \cite{Matthee2023} and \cite{Greene2024}. This correlation between \betaopt\ and \MBH\ is first found by \cite{Matthee2023} with \MBH\ $\gtrsim 10^7$ \Msun, and here is extended to \MBH\ $\sim 10^6$ \Msun\ in our sample. This is consistent with that the optical color in these systems is influenced by AGNs.

In Figure \ref{fig:fig3}-b, we compare the locations of V-shaped BLGPs and LRDs in the standard narrow-line diagnostic diagram \citep[BPT diagram,][]{Baldwin1981}.
All V-shaped BLGPs, as well as other BLGPs, exhibit high \oiii/\hb\ ratios, indicating a high level of ionization, similar to high-\itz\ broad-line AGNs.
Three of them have very low \nii/\ha\ ratios consistent with the LRD JADES 10013704, whereas the other four are more similar to typical low-\itz\ AGNs. 

To compare the Eddington ratios ($\lambda_{\rm Edd} \equiv L_{\rm bol}/L_{\rm Edd}$, where $L_{\rm Edd} = 1.26\times10^{38}\ M_{\rm BH}/M_{\odot}$) between V-shaped BLGPs and LRDs, we estimate the bolometric luminosity $L_{\rm bol}$ utilizing the empirical relation of $L_{\rm 5100} = (\frac{L_{\rm H\alpha, broad}}{5.25\times10^{42}})^{\frac{1}{1.157}}\times 10^{44}\, \rm erg\, s^{-1}$ \citep{Greene2005} and $L_{\rm bol}=9.8\ L_{\rm 5100}$ (\citealt{McLure2004}, also see \citealt{Greene2007}). 
In Figure~\ref{fig:fig3}-c, we present the $L_{\rm bol}$ vs. \MBH\ diagram. V-shaped BLGPs systematically have lower bolometric luminosities and smaller BH masses than LRDs, while most exhibit an Eddington ratio of approximately 0.1, similar to that of LRDs \citep{Greene2020, Maiolino2023, Harikane2023, Kocevski2023}.

Overall, V-shaped BLGPs exhibit similar properties to LRDs, not only in terms of broad-line widths and rest-frame UV-to-optical SEDs but also regarding rest-frame UV absolute magnitudes, the relation of optical slopes and BH masses, narrow-line diagnostics, and Eddington ratios.
These similarities make V-shaped BLGPs a valuable local sample to study the early growth and evolution of SMBHs in a local context.

\begin{figure}[htb]
    \centering
    \includegraphics[width=0.45\textwidth]{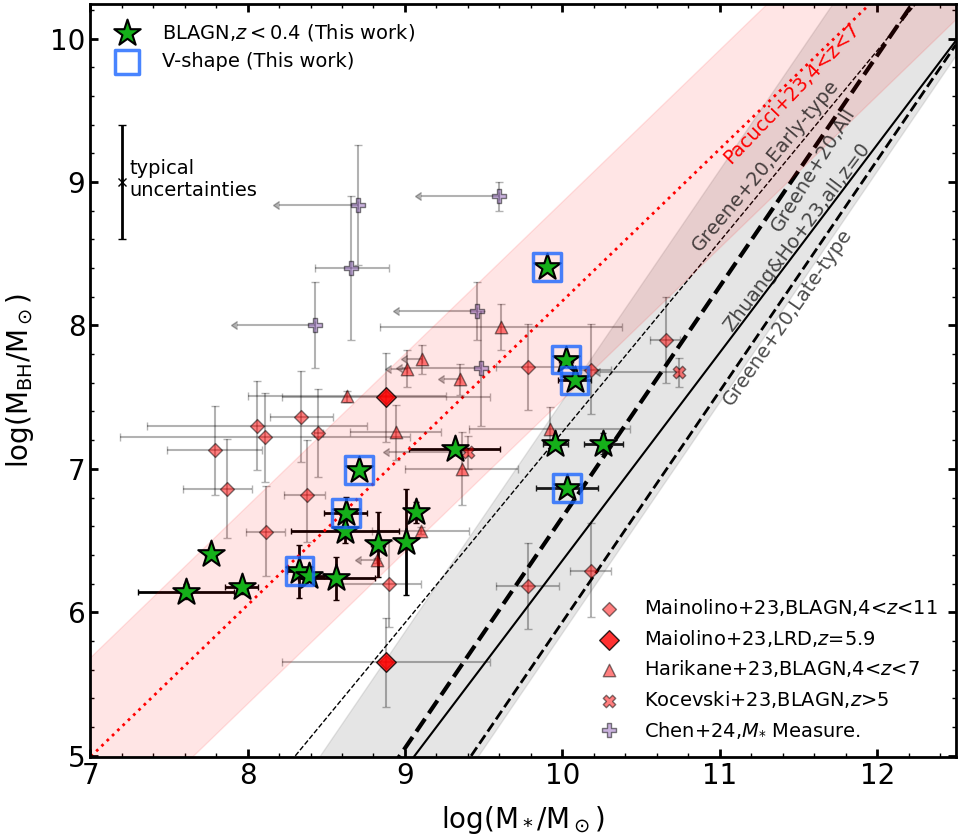}
    \caption{\MBH\ vs. \Mstar\ diagram. 
    For comparison, we illustrate high-{\it z} broad-line AGNs including LRDs reported in \cite{Maiolino2023}, \cite{Kocevski2023}, and \cite{Harikane2023}.
    Six LRDs with stellar mass measurements derived by \cite{Chen2024} are also plotted. The black dashed lines illustrate the local \MBH$-$\Mstar\ relations derived for all, early-type, and late-type galaxies \citep{Greene2020}. The solid line represents the scaling relation for AGNs evolved to \itz$=0$ \citep{Zhuang2023}. The dotted line represents the \MBH$-$\Mstar\ relation at $4<z<7$ \citep{Pacucci2023}. Shaded areas show 1$\sigma$ intrinsic scatters.}
    \label{fig:MBH_Mstar}
\end{figure}

\section{Discussion} \label{sec:discussion}
\subsection{Implication of V-shaped BLGPs}\label{sec:implication}
In this section, we discuss the implication of V-shaped BLGPs in revealing the physical origin of the UV emission of LRDs.
To distinguish between different physical origins of UV emission—whether from star formation or AGN scattered light—we can use UV emission lines (such as [Mg{\sc ii}], C{\sc iii}], and C{\sc iv}) or UV morphology. However, it remains a challenge for the high-\itz\ sample, as higher-resolution spectra or deeper imaging with JWST in the rest-frame UV band is needed \citep[e.g.,][]{Greene2024, Matthee2024}. 

In contrast, for low-\itz\ V-shaped BLGPs, obtaining the UV observations with high sensitivity and high spatial resolution is more feasible. 
With this sample of V-shaped BLGPs, we can further determine the exact physical origin of the UV emission. For instance, the most massive two V-shaped BLGPs, J120052+331238 and J092834+292136, exhibit significant Mg {\sc ii} lines (See Figure~\ref{fig:sed_spec} and Table~\ref{tab:v-shapedGP}), implying that its UV emission originates from scattered light of the center AGN.
In the future, a solid conclusion will be made by utilizing deep UV spectroscopic and imaging observations with high spatial resolution taken from instruments such as HST/COS.

\subsection{Low-\itz\ Over-massive BHs}\label{sec:overmassiveBH}
Over-massive BHs have been frequently observed in high-\itz\ broad-line AGNs and widely discussed in the literature \citep{Maiolino2023, Harikane2023, Kokorev2023, Ubler2023}.  As shown in Figure~\ref{fig:MBH_Mstar}, 16 BLGPs (including 6 V-shaped BLGPs) lay above the local \MBH-\Mstar\ relation \citep{Greene2020, Zhuang2023} more than 1$\sigma$, indicating the presence of over-massive BHs.  
These galaxies distribute similarly in the \MBH-\Mstar\ diagram with the broad-line AGNs (including LRDs) reported in literature \citep{Maiolino2023, Harikane2023, Kocevski2023}. 
Compared with the measurement of \cite{Chen2024}, who decomposed the AGN and host emission with multi-band images of LRDs and obtained clear stellar mass measurements, BLGPs have lower BH masses and lower \MBH/\Mstar\ which are up to 0.04.
Our sample of over-massive BHs at $z<0.4$ is rare and the first at such low redshifts. That challenges our current understanding of the BH formation.

The mechanisms behind the formation and evolution of these over-massive BHs remain uncertain.
In the early Universe, it involves seeding models such as Eddington or super-Eddington accretion of light seeds, or Eddington-limit accretion of heavy seeds \citep{Pacucci2023}.
In the local Universe, the formation of such BHs may be more complex due to the additional processes they undergo over time. Despite these complexities, the similarity of our BLGPs and high-\itz\ broad-line AGNs suggests that they have probably experienced a similar evolutionary track.
One possible mechanism for the formation of our sample is they are leftover black hole seeds, that they are formed from the early Universe, and have not experienced significant merging or evolution \citep{Mezcua2017}.

Alternately, BH coalescence and feedback can be another mechanism for the formation of low-\itz\ BLGPs.
Previous studies investigated over-massive BH samples at $0.4<z<3$ \citep{Mezcua2023, Mezcua2024}. They suggest that over-massive BHs may form through mergers and efficient gas accretion, with feedback processes suppressing star formation. This scenario supports the existence of over-massive BHs at $z<3$, although they are expected to evolve to align with the local scaling relation by $z\sim0$.
The presence of over-massive BHs in the local Universe may imply that center BH significantly regulates the star formation in these systems. 

We note that this phenomenon of having over-massive BHs for BLGPs is unlikely to result from the underestimation of stellar masses in their host galaxies. 
Here we calculate the stellar masses for broad-line AGNs in \cite{Maiolino2023} with our method. The median and average offsets from their results are 0.10 dex and -0.07 dex, respectively, indicating that our method has no systematic deviation from theirs.

\section{Conclusion}\label{sec:summary}
Given the similarities in physical properties between GPs and high-\itz\ SFGs, we investigated whether BLGPs could be local analogs to LRDs that also host broad-line AGNs. By comparing the rest-UV and rest-optical continuum slopes of 19 BLGPs with those of LRDs, we identified 7 BLGPs likely to be local analogs to LRDs. 

These V-shaped BLGPs exhibit faint rest-UV absolute magnitudes and sub-Eddinton accretion rates similar to those of LRDs. Three of them occupy star-formation locus in the BPT diagram, suggesting comparable ionized conditions and gas-phase metallicities similar to LRDs.
Most V-shaped BLGPs (6/7) host BHs more massive than the local scaling relation of dwarf galaxies, which is similar to LRDs. 

In general, V-shaped BLGPs and LRDs are similar populations at different redshifts, sharing similarities in rest-frame UV-to-optical SEDs, UV absolute magnitudes, narrow-line ratios, Eddington ratios, and the \MBH-\Mstar\ relation. 

Additionally, we find that most (16/19) BLGPs host over-massive BHs with masses above the local relation by more than $1\sigma$ and distribute similarly to high-\itz\ broad-line AGNs in the \MBH-\Mstar\ diagram. This over-massive BH sample is rare and the first sample at $z<0.4$, implying the challenge of understanding the BH formation.

GPs provide a local subsample that can be considered analogs to LRDs and  high-\itz\ broad-line AGNs. Combined with follow-up observation including spectroscopic and imaging observations with high sensitivity and high spatial resolution in the UV and optical bands, this V-shaped BLGP sample can help to address several unclear questions for LRDs, such as the UV emission origin.

\section*{Acknowledgments}
This work is supported by National Key R\&D Program of China No.2022YFF0503402. ZYZ acknowledges the support by the China-Chile Joint Research Fund (CCJRF No. 1906). We also acknowledge the science research grants from the China Manned Space Project, especially, NO. CMS-CSST-2021-A04, CMS-CSST-2021-A07. LCH acknowledges National Science Foundation of China (11721303, 11991052, 12011540375, 12233001), the National Key R\&D Program of China (2022YFF0503401), and the China Manned Space Project (CMS-CSST-2021-A04, CMS-CSST-2021-A06). LFB acknowledges the support of ANID BASAL project FB210003 and FONDECYT project 1230231.

Guoshoujing Telescope (the Large Sky Area Multi-Object Fiber Spectroscopic Telescope LAMOST) is a National Major Scientific Project built by the Chinese Academy of Sciences. Funding for the project has been provided by the National Development and Reform Commission. LAMOST is operated and managed by the National Astronomical Observatories, Chinese Academy of Sciences.

Funding for the Sloan Digital Sky Survey V has been provided by the Alfred P. Sloan Foundation, the Heising-Simons Foundation, the National Science Foundation, and the Participating Institutions. SDSS acknowledges support and resources from the Center for High-Performance Computing at the University of Utah. The SDSS web site is \url{www.sdss.org}. SDSS is managed by the Astrophysical Research Consortium for the Participating Institutions of the SDSS Collaboration, including the Carnegie Institution for Science, Chilean National Time Allocation Committee (CNTAC) ratified researchers, the Gotham Participation Group, Harvard University, Heidelberg University, The Johns Hopkins University, L’Ecole polytechnique f{\'e}d{\'e}rale de Lausanne (EPFL), Leibniz-Institut f{\"u}r Astrophysik Potsdam (AIP), Max-Planck-Institut f{\"u}r Astronomie (MPIA Heidelberg), Max-Planck-Institut f{\"u}r Extraterrestrische Physik (MPE), Nanjing University, National Astronomical Observatories of China (NAOC), New Mexico State University, The Ohio State University, Pennsylvania State University, Smithsonian Astrophysical Observatory, Space Telescope Science Institute (STScI), the Stellar Astrophysics Participation Group, Universidad Nacional Aut{\'o}noma de M{\'e}xico, University of Arizona, University of Colorado Boulder, University of Illinois at Urbana-Champaign, University of Toronto, University of Utah, University of Virginia, Yale University, and Yunnan University.
This research has made use of data obtained from the Chandra Source Catalog, provided by the Chandra X-ray Center (CXC) as part of the Chandra Data Archive.

%

\facilities{SDSS, LAMOST, GALEX, WISE}


\software{astropy \citep{2013A&A...558A..33A,2018AJ....156..123A}}





\bibliography{sample631}{}
\bibliographystyle{aasjournal}



\end{document}

%% file: tab_para.tex
\begin{table*}
\centering 
\caption{Parameters for 19 BLGPs with FWHM$_{\rm H\alpha}>1000\, \rm km\,s^{-1}$} 
\label{tab:parameter} 
\setlength{\tabcolsep}{1mm}{
\begin{tabular}{lccccccccc}
\hline \hline 
Name & {\it z} & \betauv & \betaopt & log $M_{\rm BH}$ & log $M_{*}$  & $M_{\rm UV}\ ^b$ & log\ \nii/\ha & log\ \oiii/\hb & dataset \\
   &    &    &    &  [\Msun] & [\Msun] & [mag] &   &   &  \\
(1) & (2) & (3) & (4) & (5) & (6) & (7) & (8) & (9) & (10)\\
\hline 
J003515+084859$^\dag$ & 0.242 & -1.24(0.71) & 0.92(0.92) & 6.99(0.02) & 8.71(0.03) & -19.14(0.32) & -1.68(0.05) & 0.78(0.01) & LAMOST/SDSS \\ 
J084029+470710 & 0.042 & -1.66(0.13) & -2.89(0.21) & 6.41(0.01) & 7.76(0.03) & -17.77(0.06) & -2.08(0.03) & 0.77(0.01) & SDSS \\ 
J085058+303053$^\dag$ & 0.280 & -1.19(0.55) & 0.44(1.18) & 6.29(0.18) & 8.33(0.08) & -19.22(0.24) & -1.33(0.17) & 0.78(0.01) & LAMOST/SDSS \\ 
J092359+415736$^\dag$ & 0.323 & -0.54(0.66) & 0.97(0.47) & 6.86(0.04) & 10.03(0.20) & -18.88(0.30) & -0.56(0.02) & 1.10(0.02) & SDSS \\ 
J092834+292136$^\dag$ & 0.293 & -0.39(0.42) & 0.21(0.31) & 7.76(0.01) & 10.02(0.04) & -19.71(0.19) & -1.24(0.03) & 0.95(0.01) & SDSS \\ 
J093818+411740 & 0.280 & -2.01(0.30) & -0.15(0.54) & 6.47(0.23) & 8.83(0.04) & -20.74(0.13) & -1.70(0.69) & 0.77(0.02) & LAMOST/SDSS \\ 
J095618+430727 & 0.276 & -1.39(0.43) & -0.22(0.52) & 6.70(0.07) & 9.07(0.02) & -20.16(0.19) & -1.14(0.07) & 0.55(0.01) & LAMOST \\ 
J112615+385817 & 0.337 & 0.21(1.09) & 0.36(0.39) & 7.18(0.07) & 10.26(0.13) & -18.29(0.47) & -0.43(0.07) & 1.18(0.05) & SDSS \\ 
J114840+175633 & 0.079 & -1.37(0.11) & -1.33(0.34) & 6.25(0.03) & 8.39(0.10) & -18.38(0.04) & -1.96(0.03) & 0.85(0.01) & SDSS \\ 
J115438+065025$^\dag$ & 0.269 & -0.45(0.44) & 0.65(0.19) & 7.62(0.03) & 10.08(0.10) & -19.78(0.20) & $<$ -1.22 & 1.02(0.05) & LAMOST \\ 
J120052+331238$^\dag$ & 0.303 & -1.85(0.98) & 1.94(0.33) & 8.41(0.03) & 9.90(0.03) & -19.42(0.45) & -0.71(0.02) & 1.01(0.01) & SDSS \\ 
J122245+360218$^\dag$ & 0.301 & -1.46(0.80) & $<$ 7.43 & 6.69(0.11) & 8.62(0.13) & -19.18(0.35) & $<$ -2.00 & 0.95(0.02) & SDSS \\ 
J124330-024241 & 0.210 & -1.10(0.78) & -4.28(2.34) & 6.57(0.08) & 8.62(0.34) & -18.43(0.36) & -1.62(0.11) & 0.77(0.01) & LAMOST \\ 
J140551+515517 & 0.271 & --$^a$ & -0.35(0.40) & 6.49(0.37) & 9.00(0.03) & --$^a$ & -0.84(0.16) & 0.79(0.06) & LAMOST/SDSS \\ 
J142644+271151 & 0.356 & -2.15(0.57) & -0.72(0.41) & 7.14(0.04) & 9.32(0.29) & -20.00(0.25) & -0.74(0.07) & 1.01(0.02) & SDSS \\ 
J160550+440540 & 0.320 & --$^a$ & 1.93(0.31) & 7.18(0.03) & 9.96(0.08) & --$^a$ & -0.73(0.03) & 1.03(0.01) & SDSS \\ 
J222803+162200 & 0.226 & -1.85(0.30) & -2.34(0.96) & 6.24(0.15) & 8.56(0.24) & -19.55(0.12) & -1.47(0.04) & 0.79(0.01) & LAMOST \\ 
J225059+000032 & 0.081 & -1.27(0.33) & -1.84(0.71) & 6.17(0.05) & 7.96(0.10) & -17.12(0.15) & -1.85(0.05) & 0.83(0.03) & SDSS \\ 
J225108-003013 & 0.081 & -0.50(0.47) & -1.63(0.48) & 6.14(0.05) & 7.61(0.31) & -16.55(0.21) & -1.89(0.04) & 0.76(0.00) & LAMOST \\ 
\hline  

\end{tabular} 
} 
    \begin{tablenotes} 
    \item NOTE. Column (1): Galaxy name. Column (2): Spectroscopic redshift. Column (3): UV continuum slope. Column (4): Optical continuum slope. Column (5): Logarithmic BH masses. Column (6) Logarithmic stellar masses. Column (7): Absolute UV magnitudes directly converted from NUV magnitudes$^b$. Column (8)-(9): Logarithmic ratio of narrow lines for \nii/\ha\ and \oiii/\hb, respectively. Column (10): spectrum dataset. The values in parentheses are 1$\sigma$ error.\\
    $^\dag$ V-shaped GPs\\
    $^a$ The galaxy is not covered by the {\it GALEX} UV observation.\\
    $^b$ As the median deviation between the $M_{\rm UV}$ converted from the NUV magnitude and $M_{1500}$ is less than 0.31 mag and negligibly impacts our result, we use the directly converted $M_{\rm UV}$.
    
    \end{tablenotes}
\end{table*}